\documentstyle[12pt,a4,epsfig]{article}
\def\rycina#1#2#3#4#5{
     \begin{figure}[htbp]
     \begin{center}
     \mbox{\epsfig{figure=#1,height=#2,width=#3}}
     \end{center}
     \caption{#4}
     \label{#5}
     \end{figure}}
\begin{document}
\vspace*{4.0cm}
\begin{large}
\begin{center}
{\Large \bf 
Experimental results  and the hypothesis 
of tachyonic neutrinos}\\ 

\vspace{1.0cm}

{ Jacek Ciborowski}\\
{Department of Physics, University of Warsaw,\\
        PL-00-681 Warsaw,  Ho\.{z}a 69, Poland}\\        

\vspace{0.4cm}

{ Jakub Rembieli\'nski}\\
{Department of Theoretical Physics, University of \L\'od\'z,\\
        PL-90-236 \L\'od\'z,  Pomorska 149/153, Poland}\\

\vspace{0.8cm}

{}
\end{center}
\end{large}
\vspace{2.5cm}

\begin{quotation}
{\bf Abstract:}\\ \sloppy
Recent measurements of the electron and muon neutrino
masses squared are interpreted as an indication that neutrinos
are faster than light particles -- tachyons. 
The tritium beta decay amplitude is calculated for the
case of the tachyonic electron neutrino. Agreement  of the theoretical
prediction  with the  shape of the recently measured  
electron spectra   is discussed.
Amplitude for the three body decay of the tachyonic neutrino,
$\nu_{i} \rightarrow \nu_{i} \nu_{j} \overline{\nu}_{j} $, 
is  calculated. It is shown that this decay may explain
the solar neutrino problem without assuming neutrino oscillations.
Predictions for short and long baseline experiments  are  commented. 
Future experimental activities are suggested.

\end{quotation}

\newpage

\section{Introduction}\label{section:introduction}

In all recent  measurements 
the values of the mass squared obtained for 
the electron \cite{backe96,lobashev96,stoeffl95,
belesev95,weinheimer93,holzschuh92,
kawakami91,robertson91} 
and muon  \cite{daum96,pdg94} neutrinos are  negative,  
as summarised  in table~\ref{table_mass2}.
This observation  requires an explanation 
which should be searched for on the grounds of both
conventional as well as  nonstandard  concepts.

At present the  best  method of determining the mass of the electron
neutrino is to fit an appropriate theoretical formula
in the endpoint region of the electron energy spectrum
measured in the $\beta$  decay  of tritium
(electron kinetic energy at endpoint amounts to $\approx 18570\;{\rm eV}$).

\begin{table}[htb]
\begin{center}
\begin{tabular}{|c|rclr|c|c|}
\hline
 Flavour  &  &$m^{2}$& && Year  & Ref. \\
\hline
      & $-22$&$\pm$& $17\pm 14\;$ &${\rm eV}^{2}$&1996&\cite{backe96}  \\
      & $-20.6$&$\pm$& $5.8\;$ &${\rm eV}^{2}$&1996&\cite{lobashev96}  \\
      & $-130$&$\pm$& $20\pm 15\;$ &${\rm eV}^{2}$ & 1995 & \cite{stoeffl95}  \\
 $\nu_{e}$  & $-22$&$\pm$& $4.8 \;$ &${\rm eV}^{2}$ & 1995 & \cite{belesev95}  \\
      & $-39$&$\pm$& $34\pm 15\;$ &${\rm eV}^{2}$&1993&\cite{weinheimer93}  \\
      & $-24$&$\pm$& $48\pm 61\;$ &${\rm eV}^{2}$&1992&\cite{holzschuh92}  \\ 
      & $-65$&$\pm$& $85\pm 65\;$ &${\rm eV}^{2}$&1991&\cite{kawakami91}  \\
      &$-147$&$\pm$& $68\pm 41\;$ &${\rm eV}^{2}$&1991&\cite{robertson91}  \\
\hline
 $\nu_{\mu}$  &  $-0.143$&$\pm$ &$0.024\;$ &${\rm MeV}^{2}$&1996&\cite{daum96}  \\
    &  $-0.016$&$\pm$& $0.023\;$ &${\rm MeV}^{2}$&&  \\
\hline
\end{tabular}
\caption{Recent results of the electron and muon neutrino mass squared 
measurements. The first uncertainty is statistical, the second systematic.}
\label{table_mass2}
\end{center}
\end{table}

Improving  resolution  of spectrometers used in certain  recent  experiments 
lead to an  observation of two unexpected details 
in  the  electron energy spectrum,  
presently often referred to as    {\it anomalies}.

One {\it anomaly}  appears  as an  enhancement   
located  close to the endpoint  in the integral  
electron energy spectrum \cite{lobashev96,stoeffl95,belesev95}, 
as shown in figs.~\ref{ryc:stoeffl95}  and  ~\ref{ryc:belesev95}.    
Branching ratio 
of this structure comes out to be  of the order 
of     $3\cdot 10^{-9}$ \cite{stoeffl95}   or  $10^{-10}$ \cite{belesev95}.
Presently an effort is spent to 
understand the effect on the grounds of atomic physics \cite{workshop96}
and statistics \cite{khalfin96}.

\rycina{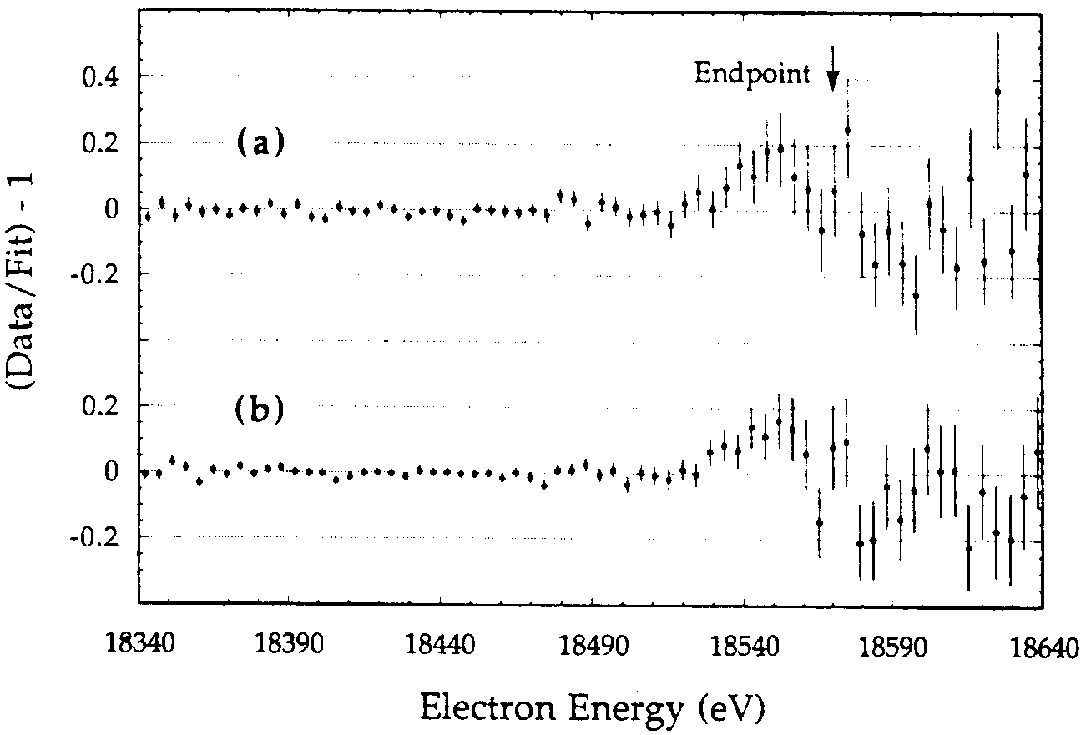}{8cm}{8cm}
{{\sl Ratio of measured and fitted spectra (Kurie plot)  
from  the LLNL  experiment \protect{\cite{stoeffl95}}. 
}}
{ryc:stoeffl95}

\rycina{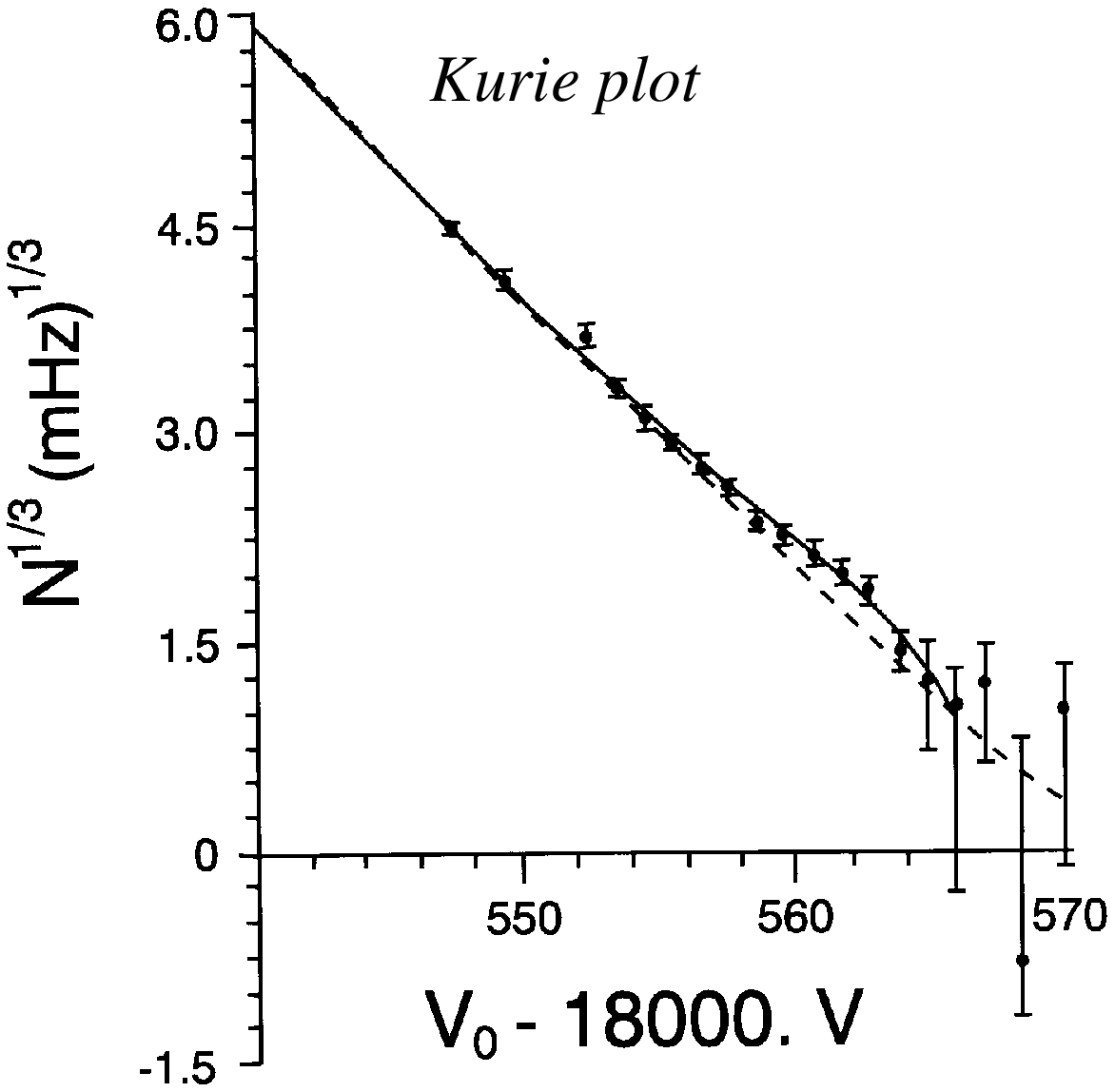}{8cm}{8cm}
{{\sl Integral electron energy spectrum in tritium decay (Kurie plot) 
measured in the Troitsk experiment \protect{\cite{belesev95}}.  
}}
{ryc:belesev95}

The second {\it anomaly}, located  at lower  energies,
is related to the slope of the measured 
electron energy spectrum. If the  theoretical  spectrum
is fitted  in the energy range below approximately $18300\; {\rm eV}$
then the endpoint  occurs  at a lower value of energy.
This {\it anomaly} was interpreted as a manifestation of 
a missing spectral component   with about
$4\%$ branching ratio and the endpoint energy  around $18400\; {\rm eV}$ \cite{belesev95}
or $18500\; {\rm eV}$ \cite{weinheimer93}.

The mass squared  of the muon neutrino has been determined from the 
measured muon momentum in the  $\pi^{+}$ decay at rest.  
The latest, most precise  results     also indicate a negative
value \cite{daum96}. The two entries in table~\ref{table_mass2}
correspond to different values of the charged pion mass \cite{pdg94}. 
This ambiguity soon should be resolved by additional  measurements 
\cite{daum96}.

Until now there have been no experiments dedicated  to  determine the
tau neutrino mass. The latest analysis  \cite{aleph_tau_nu} lead to
a result with large uncertainties and as such 
cannot be considered in the context of the idea presented
in this paper.

In view of the above   results for the  neutrino masses squared 
we  elaborate on the hypothesis that   neutrinos are  tachyons. 
The aim of this paper is to present  a confrontation  of  experimental data
with theoretical calculations performed for processes involving 
tachyonic neutrinos. 
It will be shown that the measured shape of the electron spectrum
near the endpoint in the  tritium decay  is qualitatively consistent
with that predicted from  the decay  involving 
the  tachyonic electron neutrino.
The amplitude for the three body lepton flavour conserving decay of the
tachyonic neutrino,
$\nu_{i} \rightarrow \nu_{i} \nu_{j} \overline{\nu}_{j} $, 
is  calculated and  predictions for certain phenomena discussed.

In the following sections various  values of  tachyonic
neutrino masses appear as a result of  certain assumptions.
Tachyons, like massive and massless particles,  
are subject to  gravitational interactions too.  
However at  the present,  early   stage of our  work 
we have not yet grounds to comment possible  constraints 
on tachyonic neutrino masses resulting from  considerations
on the cosmological scale.

\section{Theoretical considerations}\label{section:theor}

It has been a common conviction that  tachyons could not exist for  several 
reasons related to fact that they are  moving with superluminal
velocities:  negative energies, causality violation, vacuum instability
and others. Despite a significant theoretical effort no solution
of these problems has been  found within the framework of the 
Einstein--Poincar\'e  (EP) relativity. 
However the  recently   proposed  causal theory of tachyons 
\cite{jaremb1,jaremb2,jaremb3}  is free of these difficulties.

The main idea is based on two well known facts: $(i)$ the definition of
a coordinate time depends on the synchronisation scheme; $(ii)$ 
synchronisation scheme is a convention, because no experimental procedure
exists which makes it possible to determine the one-way velocity
of light without use of superluminal signals. Therefore there is a freedom
in the definition of the coordinate time. The standard choice is the
Einstein--Poincar\'e (EP) synchronisation with one-way light velocity
isotropic and constant. This choice leads to the extremely simple form of
the Lorentz group transformations but the EP coordinate time implies
a covariant causality for time-like and light-like trajectories only.
We choose a  different synchronisation, namely that  of 
Chang--Tangherlini (CT) preserving invariance of the notion of the
instant-time hyperplane \cite{ct1,ct2}. 
In this synchronisation scheme the
causality notion is universal and space-like trajectories are
physically admissible too. The price is the more complicated form
of the Lorentz transformations incorporating transformation rules
for velocity of distinguished reference frame (preferred frame).
{\bf
The EP and CT descriptions are entirely equivalent if we restrict
ourselves to time-like and light-like trajectories;
however a consistent description of tachyons is possible only
in the CT scheme.}
A very important consequence is that if tachyons exist then the relativity
principle is broken,  i.e. there exists a preferred frame of reference,  
however the Lorentz symmetry is   preserved.
As we know, in the real world
we have such a  locally distinguished frame -- namely the cosmic background
radiation frame.

The interrelation between EP ($x_E$) and CT ($x$)coordinates reads:
\begin{equation}
x_{E}^{0}=x^0+u^0 \vec{u} \vec{x} \quad,\qquad \vec{x}_E=\vec{x}
\label{eq:theor1}
\end{equation}
where $u^{\mu}$ is the four-velocity of the privileged frame as seen
from the frame ($x^{\mu}$).

In  refs.~\cite{jaremb1,jaremb2,jaremb3} a  fully consistent,  Poincar\'e
covariant quantum field theory of  tachyons was given. 
In particular,  the elementary
tachyonic states are labelled by helicity. In the case of the fermionic
tachyon with helicity $-\frac{1}{2}$ the corresponding field equation
reads:
\begin{equation}
\left(\gamma^5\left(i\gamma\partial\right)-\kappa\right)\psi=0
\label{eq:theor2}
\end{equation}
where the bispinor field $\psi$ is simultaneously an eigenvector of the
helicity operator with eigenvalue $\frac{1}{2}$. Here the $\gamma$-matrices
are expressed by the standard ones in analogy to the relation 
(\ref{eq:theor1}).
The solution of (\ref{eq:theor2}) is given by:
\begin{equation}
\psi(x, u) \\ = \frac{1}{(2\pi)^{\frac{3}{2}}}\int d^4k\,\delta(k^2+\kappa^2)
\theta(k^0)\left[w(k,u)e^{ikx}b^\dagger(k)
+v(k,u)e^{-ikx}a(k)\right] \label{eq:theor3}
\end{equation}
where the operators $a$ and $b$ correspond to neutrino and antineutrino,
respectively. The amplitudes $v$ and $w$ satisfy the following
conditions:
\begin{equation}\label{eq:w-w}
w(k, u) \bar{w}(k, u) = (\kappa-\gamma^5k\gamma)
\frac{1}{2}\left(1-\frac{\gamma^5 [k \gamma, u \gamma]}{2
\sqrt{q^2+\kappa^2}}\right)
\end{equation}

\begin{equation}\label{eq:v-v}
v(k, u) \bar{v}(k, u) = -(\kappa+\gamma^5k\gamma)
\frac{1}{2}\left(1-\frac{\gamma^5 [k \gamma, u \gamma]}{2
\sqrt{q^2+\kappa^2}}\right)
\end{equation}

\begin{equation}\label{eq:n1}
\bar w(k,u)\gamma^5 u \gamma w(k,u)
=\bar v(k,u)\gamma^5 u \gamma v(k,u)=2q
\end{equation}

\begin{equation}\label{eq:n2}
\bar w(k^{\Pi},u)\gamma^5 u \gamma v(k,u)=0.
\end{equation} 

\noindent
Here $q=u_{\mu}k^{\mu}$ is equal to the energy of the tachyon in the
preferred frame and  $\Pi$ denotes the space 
inversion operation.
It is easy to check that in the massless limit
$\kappa\rightarrow 0$ the above relations give the Weyl's theory.

\section{Tritium $\beta$ decay with  the tachyonic neutrino}  

Until now the  mass squared of the electron neutrino has been determined 
by fitting the measured integral electron energy spectrum with 
a function comprising an  expression  for  the decay probability.
The integral  form of this expression is as follows
(neglecting summation over final states) \cite{weinheimer93}:

\begin{equation} \label{se}
  S(E) \sim  p_{e}(E_{e}+m_{e})
(E_{0}-E_{e})\sqrt{(E_{0}-E_{e})^{2}-m_{\nu_{e}}^{2}}
\end{equation}

\noindent
where $E_{e},\;p_{e},\;m_{e}$  are the electron kinetic energy, momentum
and mass, $E_{0}$ is the  endpoint  kinetic energy
and $m_{\nu_{e}}^{2}$ is the square of the electron antineutrino mass.
This formula is obtained from  the standard
calculation of  the  three body weak decay, including a massive
neutrino.  
However if one wants to test the hypothesis   that  neutrinos are tachyons, 
this formula is no longer valid for that purpose even if the 
sign in front of the  $ m_{\nu_{e}}^{2} $   term is changed.  
The correct expression can be obtained
from a similar, although by far more complicated,
calculation.

Let us consider the $\beta$ decay in the  
framework of section~\ref{section:theor},
using effective four-fermion interaction. In the first order of
the perturbation series the decay rate for the process:
$n\longrightarrow p^{+} + e^{-} + \overline{\nu}_{e}$ with the tachyonic
electron antineutrino reads:
\begin{equation}
d\Gamma=\frac{1}{4 m_{n} (2\pi)^5}d\Phi_3 \left|M\right|^2,
\label{eq:beta1}
\end{equation}
where $d\Phi_3$ is the phase-space volume element:
\begin{eqnarray}
\lefteqn{d\Phi_3=\,\theta(k^0)\,\theta(l^0)\,\theta(r^0)}\cdot\nonumber\\
&&\delta(k^2-m_{p}^{2})\,\delta(l^2-m_{e}^{2})\,\delta(r^2+\kappa^2)\,
\delta^4(p-k-l-r)\,d^4k\, d^4l\, d^4r
\label{eq:beta2}
\end{eqnarray}
while
\noindent
\begin{equation}
\left|M\right|^2=2 \,G_{F}^{2}\,
{\rm Tr}\left[u_e \overline{u}_e \gamma^{\mu} w \overline{w}
\gamma^{\nu}\right] \\  
{\rm Tr}\left[u_p \overline{u}_{p}
\gamma_{\mu}\left(1-g_A\gamma^5\right) u_n \overline{u}_n
\gamma_{\nu}\left(1-g_A\gamma^5\right)\right].
\label{eq:beta3}
\end{equation}
Here $p,k,l,r$ are the four-momenta of $n,p^+,e^-$ and
$\overline{\nu}$ respectively; the corresponding masses are denoted
by $m_n,m_p,m_e$ and $\kappa$
(in the case of tritium decay $m_{n}$ and  $\;m_{p}$ 
denote   the masses  of $^{3}H$ and $^{3}He$). 
$G_F$ and $g_A$ are the
Fermi constant and the axial coupling constant. The amplitudes
$u_n,\overline{u}_n,u_p,\overline{u}_p,u_e,\overline{u}_e$
satisfy usual\footnote{but with $\gamma$ in the CT synchronisation}
polarisation relations:
$u_n \overline{u}_n=p\gamma+m_n$,
$u_p \overline{u}_p=k\gamma+m_p$,
$u_e \overline{u}_e=l\gamma+m_e$
whereas
$w \overline{w}$ is given by   eq.~\ref{eq:w-w}.
After elementary calculations eq.~\ref{eq:beta3} reads:

\begin{eqnarray}
\lefteqn{\left|M\right|^2=16\,G_{F}^{2}\left\{\left[m_n m_p (1-g_{A}^{2})
-(kp)(1+g_{A}^{2})\right]\right.\cdot}\nonumber\\
&& \cdot\left(4 m_e \kappa - \frac{1}{\sqrt{(ur)^2+\kappa^2}}
\left[4 \left(\kappa^2(lu)+(lr)(ur)\right)-2\kappa^2(lu)-
2(ur)(lr)\right]\right)+\mbox{}\nonumber\\
&&\mbox{}+(1+g_{A}^{2})\left(2 m_e \kappa (pk) -
\frac{1}{\sqrt{(ur)^2+\kappa^2}}\right.\cdot\nonumber\\
&&\mbox\quad\cdot\left[2 (pk)
\left(\kappa^2(lu)+(lr)(ur)\right) -
2 \kappa^2 \left((pl)(uk)+(kl)(up)\right)\right.+\mbox{}\nonumber\\
&&\mbox{}\qquad - \left.\left.2 (ur) \left((pl)(kr)+(kl)(pr)
\right)\right]\right)+\mbox{}\nonumber\\
&&\mbox{}+4 g_A \left.\left[(pr)(kl)-(pl)(kr)+
\frac{m_e \kappa}{\sqrt{(ur)^2+\kappa^2}}
\left((kr)(up)-(pr)(uk)\right)\right]\right\}
\label{eq:beta4}
\end{eqnarray}

Because the Solar System is almost at
rest relatively to the cosmic background radiation
frame\footnote{relative velocity is about $370 {\rm km}/{\rm s}$}
we can calculate the electron energy spectrum
$d\Gamma /dl^0$ in this frame
with  sufficient precision. The resulting formula is rather
complicated (\ref{eq:dGdE1}); we present it  in the Appendix.

\begin{figure}[htb]
     \centerline{\mbox{
     \psfig{file=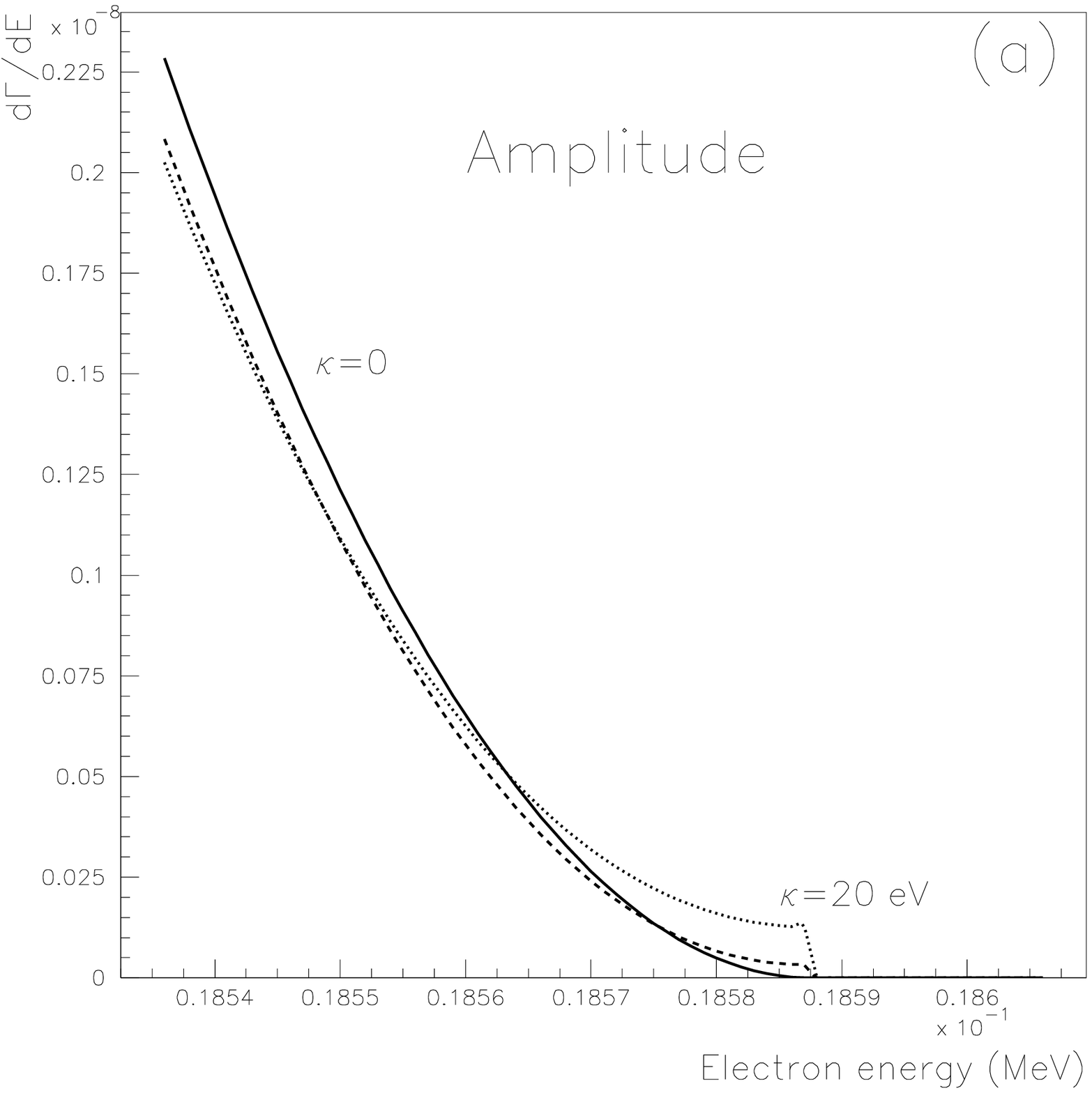,height=9cm,width=9cm}
     \psfig{file=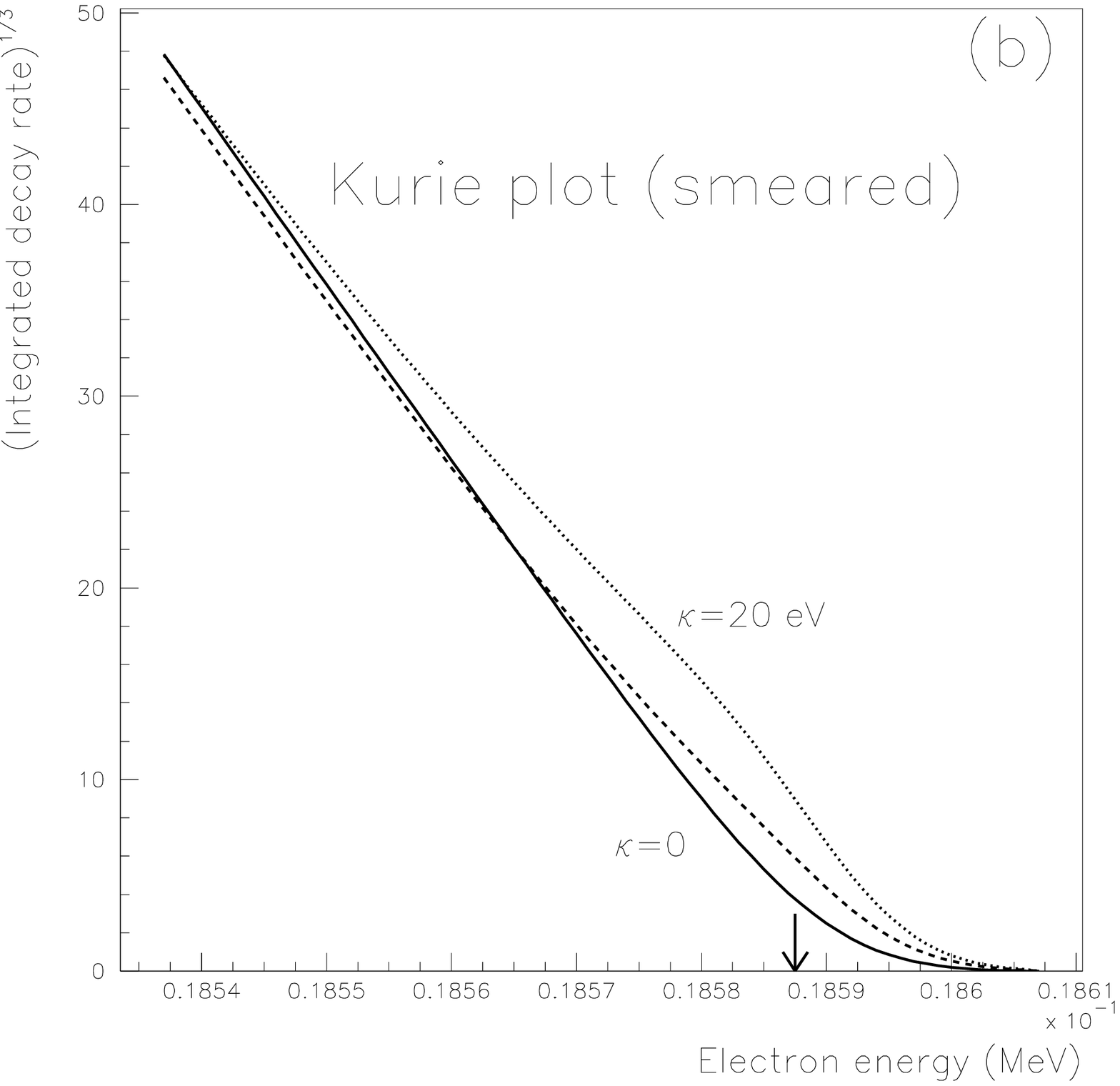,height=9cm,width=9cm}}}
  \caption{{\sl (a) Theoretical differential energy spectrum 
$d\Gamma / dE$ for  tritium
decay with the tachyonic electron antineutrino of mass $\kappa$;
(b) smeared integral spectrum  (Kurie plot) 
-- endpoint energy indicated by arrow.
Curves drawn for masses: $\kappa=0,10,20\;{\rm eV}$.
  }}
     \label{ryc:1stpicture}
     \end{figure}

Formula  (\ref{eq:beta1}) gives  identical 
results in the  limit  $\kappa^{2}\rightarrow 0$
as that for the massless neutrino.
Differential electron energy spectra near the endpoint for 
massless and tachyonic neutrinos are shown 
in fig.~\ref{ryc:1stpicture}a
(we  used  the following  values for the 
masses of $^{3}{\rm H}$ and $^{3}{\rm He}$: $2809.94\; {\rm MeV}$ and $2809.41\; {\rm MeV}$).
The distinctive feature of the spectrum near the endpoint 
in the tachyonic case is the enhancement with a step-like
termination. 
Function (\ref{eq:dGdE1}) goes to zero over a narrow range
of energy,  contrary to  a monotonic  character  in the cases 
of the massless (and massive) neutrino.
In order to allow for a qualitative comparison  with the measurement 
of ref.~\cite{belesev95}, the  electron  spectrum given
by (\ref{eq:dGdE1}) was integrated and 
smeared with  constant experimental energy resolution of $4\; {\rm eV}$
(assumed Gaussian). 

As a result  the step-like endpoint structure of the differential spectrum
appeared  as an enhancement in the  integral electron energy 
distribution, as shown in fig.~\ref{ryc:1stpicture}b. 
Position, magnitude and  width of the enhancement depend on 
the tachyonic electron neutrino mass, $\kappa$.
At this stage of investigations  
we may conclude that   using formula (\ref{eq:dGdE1})  it is possible  to
reproduce  qualitatively the enhancement in  the 
electron energy spectrum near the endpoint  if  a sufficiently large
value of the tachyonic neutrino mass is taken. 
Indeed, no `{\it bump}' is seen
for $\kappa=10\;{\rm eV}$ while it is already clearly marked for $\kappa=20\;{\rm eV}$. 
It is indispensable  to  perform a fit  to the experimental  spectrum; 
however we were not in possession neither of the data
nor the details of the apparatus. 

For the purpose of considerations presented in 
section~\ref{section:3bdecay} we need an input value for 
the tachyonic electron antineutrino mass. The value $\kappa\approx 5\;{\rm eV}$,
quoted in refs.~\cite{lobashev96,belesev95}, is  not
meaningful  for the tachyonic hypothesis
since it was  obtained by fitting   function  (\ref{se})
which describes the  massive neutrino ($m^{2}>0$). 
As an `educated guess'  (fig.~\ref{ryc:1stpicture}b) we  take 
$20\;{\rm eV}$  for the mass of the tachyonic electron neutrino.

\section{Decays  of tachyonic neutrinos}
\label{section:3bdecay}

\subsection{Introduction}

\begin{figure}[htb]
     \centerline{\mbox{
     \psfig{file=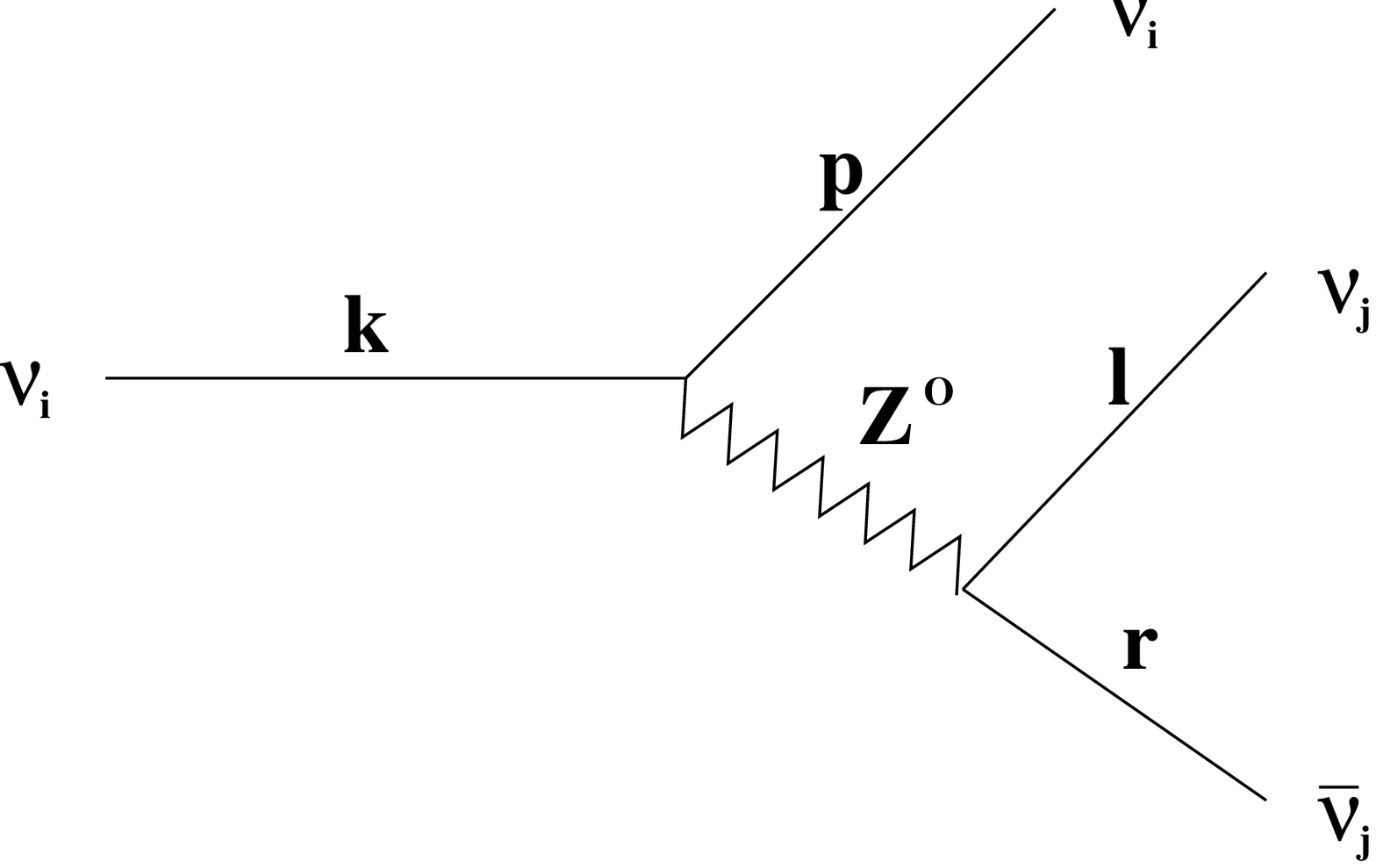,width=5cm,angle=-90} \hspace{3.0cm}
     \psfig{file=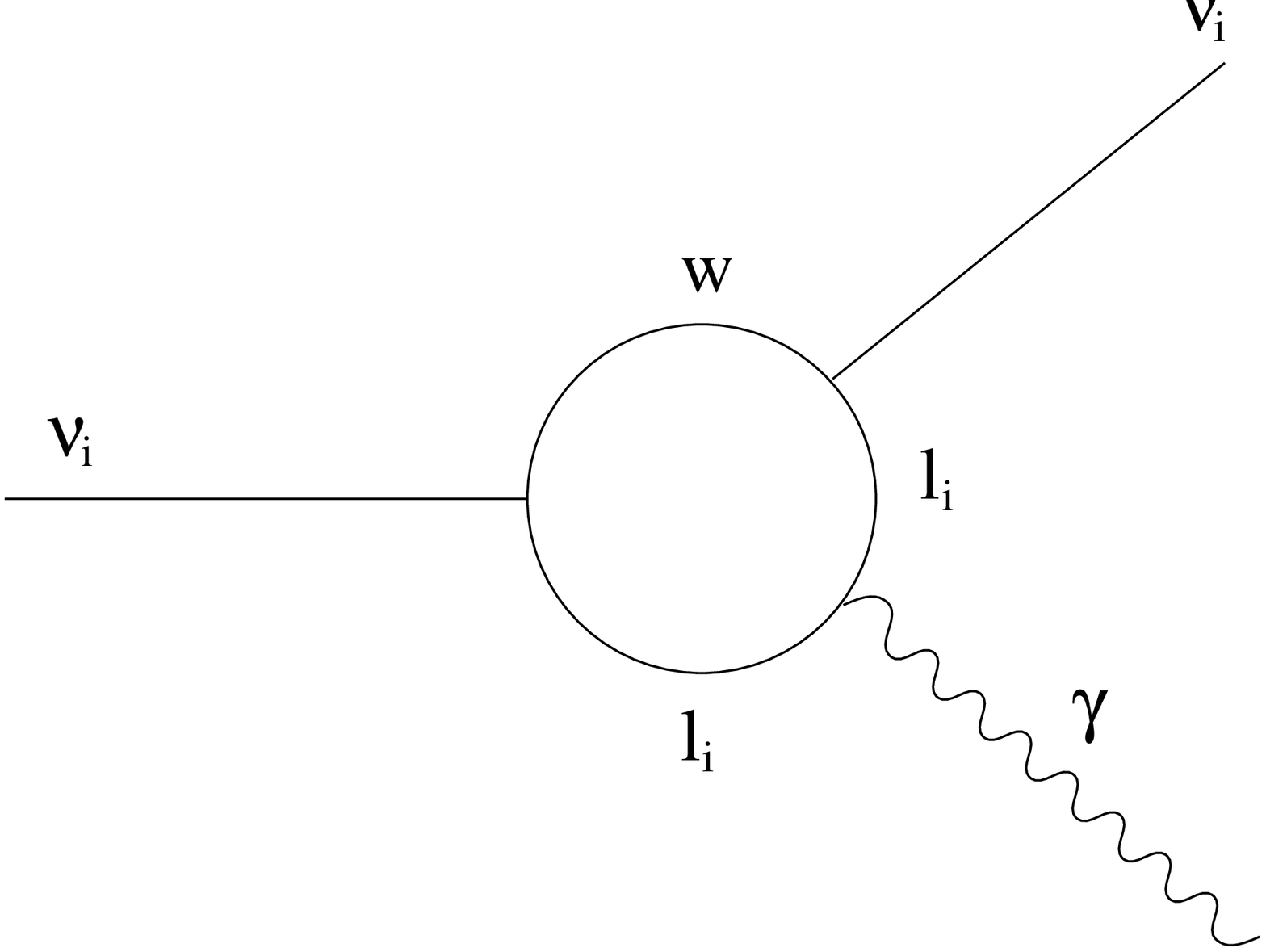,width=5cm,angle=-90}}}
     \caption{{\sl Lepton flavour conserving  decays of 
       the tachyonic neutrino.  }}
     \label{ryc:decays_graph}
     \end{figure}

Tachyonic neutrinos are in general unstable.
Similarly to a  three body decay of a  massive particle,
a tachyon may decay into three tachyons. In the latter case however
a specific channel is allowed with the initial tachyon appearing
also in the final state.  In the case of  tachyonic neutrinos
the corresponding process would be 
the  decay of a neutrino of flavour `$i$'  into itself and 
a neutrino -- antineutrino pair of flavour `$j$': 
$\nu_{i} \rightarrow \nu_{i} \nu_{j} \overline{\nu}_{j} $,
where $i,j=e,\mu,\tau$,
as shown in fig.~\ref{ryc:decays_graph}. 

Another important mode is the radiative decay, 
$\nu_{i}\rightarrow\nu_{i}\gamma$,
shown in the same figure. In both processes the lepton family number
is conserved, contrary to  those involving  a massive neutrino.
In this section we deal with the three body decay and 
calculate  partial widths $\Gamma_{ij}$ in order to find out
whether  the tachyonic hypothesis 
is not in contradiction with established observations.

Tachyonic neutrino decays  mean an additional, qualitatively similar  
process to oscillations. 
Consider a beam of neutrinos of a given flavour, characterised by  
a certain energy spectrum.  As a result of decays 
new flavours appear as a function of distance
(including antineutrinos in neutrino beams and vice versa)
but in  a pattern  much more complicated  than in the case of oscillations.
>From the theoretical point of view 
the decay has  a clear advantage over  the latter
since its magnitude can be calculated  within the 
theory of weak interactions; masses of tachyonic neutrinos, which 
can be determined from independent measurements,
are  the only  input parameters in that case.

\subsection{Derivation of formulae}

Assuming universality of weak interactions we can calculate the decay
rate for the process  
$\nu_i\rightarrow\nu_{i} \nu_j\overline{\nu}_{j}$
($i,j=e,\mu,\tau$) which is given by:
\begin{equation}\label{eq:3b1}
d\Gamma_{ij}=\frac{1}{(2\pi)^52q}d\Phi\left|T\right|^2,
\end{equation}
where
\begin{equation}\label{eq:3b2}
d\Phi=\theta(p^0)\theta(l^0)\theta(r^0)\delta(p^2+\kappa^2)
\delta(l^2+\mu^2)\delta(r^2+\mu^2)\delta(k-p-l-r)d^4pd^4ld^4r
\end{equation}
is the phase space volume element, the corresponding four-momenta $k,p,l,r$
as shown in fig.~\ref{ryc:decays_graph}, $\kappa$ and $\mu$  are the masses
of the $i$-th and $j$-th neutrino, $q$ -- energy of the $i$-th neutrino 
and
\begin{equation}\label{eq:3b3}
\left|T\right|^2=4 G_{F}^{2} {\rm Tr}\left[v_i(k)\bar{v}_i(k)
\gamma^{\mu}v_i(p)\bar{v}_i(p)\gamma^{\nu}\right]
{\rm Tr}\left[w_j(r)\bar{w}_j(r)
\gamma_{\mu}v_j(l)\bar{v}_j(l)\gamma_{\nu}\right]
\end{equation}
can be calculated using results of section~\ref{section:theor}. 
The final formula for $\Gamma(q,\kappa,\mu)$ is  rather long 
and complicated and  for this reason we do not present it  here.

\subsection{Results and discussion}

Existence of  the three body decay  raises   questions
about stability of tachyonic neutrinos in different environments,
ranging from  laboratory to the Universe.
As shown above, the mean lifetime of the tachyonic neutrino 
of a given flavour   can be calculated only if   masses 
of all three  neutrino species are known. 
The present situation in this respect may be viewed as such that the mass
of the tachyonic electron neutrino may be considered to be approximately known 
($\kappa\approx 20\;{\rm eV}$) but not those of the muon and tau neutrinos.
Uncartain  measurements  in  the two latter  cases are the principal 
difficulty in discussing the consequences of neutrino
decays in detail.  For that reason the following remarks
are to some  extent of a speculative nature.
In order to simplify the  discussion, 
we  limit our considerations  to a two flavour  scenario with the electron, 
$\nu_{e}$, and  the heavy, $\nu_{H}$, neutrinos,  implicitly assuming
$\kappa_{e}\ll\kappa_{H}$. 

\begin{figure}[htb]
     \centerline{\mbox{
     \psfig{file=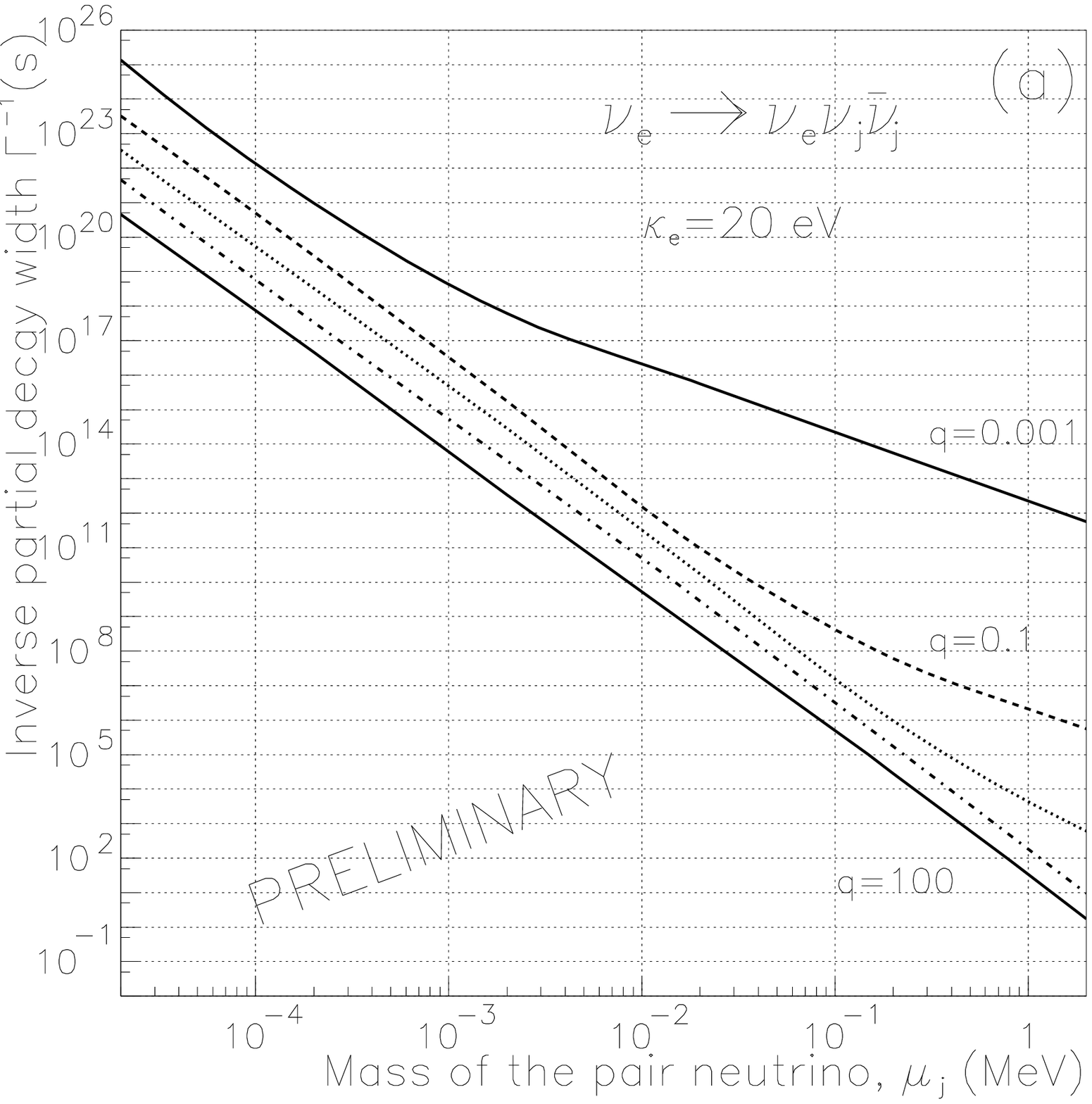,height=9cm,width=9cm}
     \psfig{file=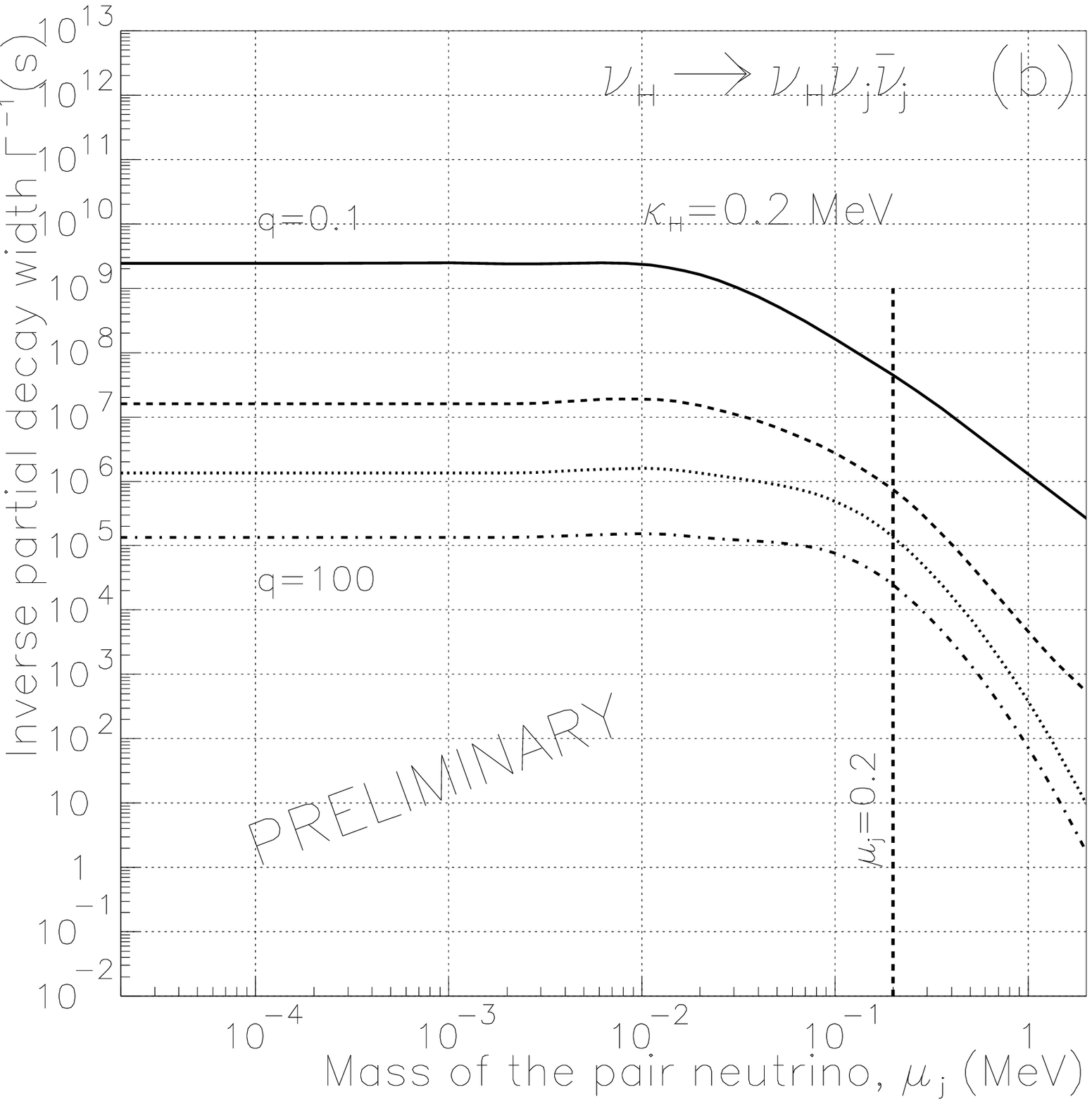,height=9cm,width=9cm}}}
  \caption{{\sl  Dependence of the inverse partial decay width 
(seconds) on $\mu_{j}$
(preliminary results  with accuracy of  few percent  due to numerical
complexity of  calculations). Curves correspond to different  decaying 
neutrino energies, $q$: $0.001$, $0.1$, $1$, 10 and $100\;{\rm MeV}$, 
in the indicated order;
(a) for the electron neutrino, $\kappa=20\;{\rm eV}$;
(b) for the heavy  neutrino, $\kappa=0.2\;{\rm MeV}$.
}}
     \label{ryc:3b_fig12}
     \end{figure}

It is a specific feature of tachyonic neutrinos that  mean
lifetime for the  three body decay decreases as a power of  energy, $q$.
Magnitude of this  dependence  varies  significantly over a wide range of 
tachyonic neutrino masses, as can be seen in fig.~\ref{ryc:3b_fig12}.
In general the process of neutrino energy degradation 
is slowing down with subsequent decays.

Partial width for the  decay of a  tachyonic neutrino into a given 
final state  increases with  the mass of the  neutrino from the pair, 
$\mu_{j}$.
For a light neutrino like $\nu_{e}$ this dependence is very strong
in a wide range of $\mu_{j}$ masses,
as can be seen from fig.~\ref{ryc:3b_fig12}a.
Total width for the electron neutrino decay
($i=e,\:j=e,H$), neglecting the radiative
decay,  is approximately equal to 
the sum of partial widths for the two channels:
$\Gamma_{e}=\Gamma_{ee}+\Gamma_{eH}$. If $\kappa_{H}\ll\kappa_{e}$
then $\Gamma_{ee}\gg\Gamma_{eH}$  so  to a good accuracy 
$\Gamma_{e}\approx\Gamma_{eH}$ which means that the mean lifetime 
of the tachyonic electron neutrino is determined by the (unknown)
mass of the heavy neutrino.
Concerning the latter, the present  measurements of the
muon neutrino mass squared  justify considering the value 
$\kappa_{H}\approx0.2\;{\rm MeV}$ which lies within a $1\sigma$  uncertainty
of the latest measurement. A heavy neutrino of this mass 
would be significantly less stable than the electron neutrino, 
as can be seen  from  fig.~\ref{ryc:3b_fig12}b;
moreover, its mean lifetime  does not depend on the exact value
of the  electron neutrino mass.
The ratio of partial widths for both decay channels, $\Gamma_{He}/\Gamma_{HH}$,
amounts to 0.1 for $q=100\;{\rm MeV}$,  making the 
$\nu_{H}\rightarrow\nu_{H}\nu_{e}\overline{\nu}_{e}$ 
decay channel not negligible.

In general, inverse partial widths may be arbitrarily large provided tachyonic
neutrino masses are sufficiently small. If the mass of the
heavy neutrino were smaller than about $100\;{\rm eV}$ then the mean lifetime
of the electron neutrino would exceed the age of the Universe for 
energies up to those presently reached in accelerator beams.
On the other hand it is interesting to speculate about consequences 
of larger  masses. 
If the mass of the heavy neutrino were of the order of $1\;{\rm MeV}$
then the mean lifetime of  
$q\approx 10\;{\rm MeV}$  solar electron neutrinos would be comparable
to the time of flight from the Sun to Earth (fig.~\ref{ryc:3b_fig12}a).
Under this condition one would expect   
both flavour composition and energy spectrum in the solar flux
to be modified due to decays. At a first glance this is an appealing
possibility to explain the  solar neutrino problem
in a simple way, without involving the phenomenon of  neutrino oscillations.
However if neutrinos of  similar (or higher)  energies 
travel a larger distance, 
like e.g. from SN1987A to Earth, their energy spectrum  would be  modified too.
We have not yet carried out a detailed co-analysis of the data from
both sources.
Under certain assumptions  we may  put an upper
limit on the heavy neutrino mass using  the SN1987A data. 
For that purpose  we assume tentatively 
that the  energy spectrum of the SN1987A  
neutrinos, measured on Earth,
is little or not at all distorted as compared to  the one 
during emission. This may be achieved if:
($i$) either a negligible fraction of neutrinos decayed
during $t_{0}\approx 5\cdot10^{12}\;s$ (150 000 years) on the way to Earth
(if $\tau\ll t_{0}$)  or ($ii$) after each decay the $\nu_{j} \overline{\nu}_{j}$
pair acquired a negligible fraction of the initial neutrino  energy
(if $\tau<t_{0}$).
Inequality  $\tau\ll t_{0}$ implies   the mass of the heavy
tachyonic neutrino  to  be much smaller than about $5\;{\rm keV}$, as can
be read from  fig.~\ref{ryc:3b_fig12}a for $q=0.1\div 10\;{\rm MeV}$.
This result should be taken with caution since there is no firm
experimental evidence for the underlying assumption. 

Three body decays of tachyonic neutrinos may simulate oscillations
in terms of experimental signature in an   experiment located at 
a fixed  distance from the source.
In order to distinguish between   oscillations and  decays  
measurements should be performed at different distances.
Candidate events for neutrino oscillation have been recently reported
in the $\overline{\nu}_{\mu} \rightarrow \overline{\nu}_{e}$ channel
at LAMPF;
if interpreted  as such, this measurement
indicates oscillation probability of 0.3\% 
(over a distance of  $\approx 30\;m$) \cite{lampf95,lampf96}. 
In fig.~\ref{ryc:3b_fig3} we show the inverse partial decay width
for the decay 
$\nu_{H} \rightarrow \nu_{H} \nu_{e} \overline{\nu}_{e} $
as a function of the decaying  neutrino mass, $\kappa_{H}$,
for energy $q=50\;{\rm MeV}$.
If we identify $\nu_{\mu}$ with $\nu_{H}$  then it follows from
the experimental conditions that $1/\Gamma$ for this decay 
channel  should be of the order 
of $10^{-4}\;s$ to observe  the above fraction of decays. 
Such a low value can only be achieved  for 
the  muon neutrino mass in the range  of dozens  ${\rm MeV}$, which is
excluded by measurements.
Thus the LAMPF  candidate events  cannot be explained by the three body decay 
of the tachyonic muon neutrino: 
$\nu_{\mu} \rightarrow \nu_{\mu} \nu_{e} \overline{\nu}_{e} $
(as can be seen in the same figure, the result does  not depend on the exact value
of the electron neutrino mass).

  \rycina{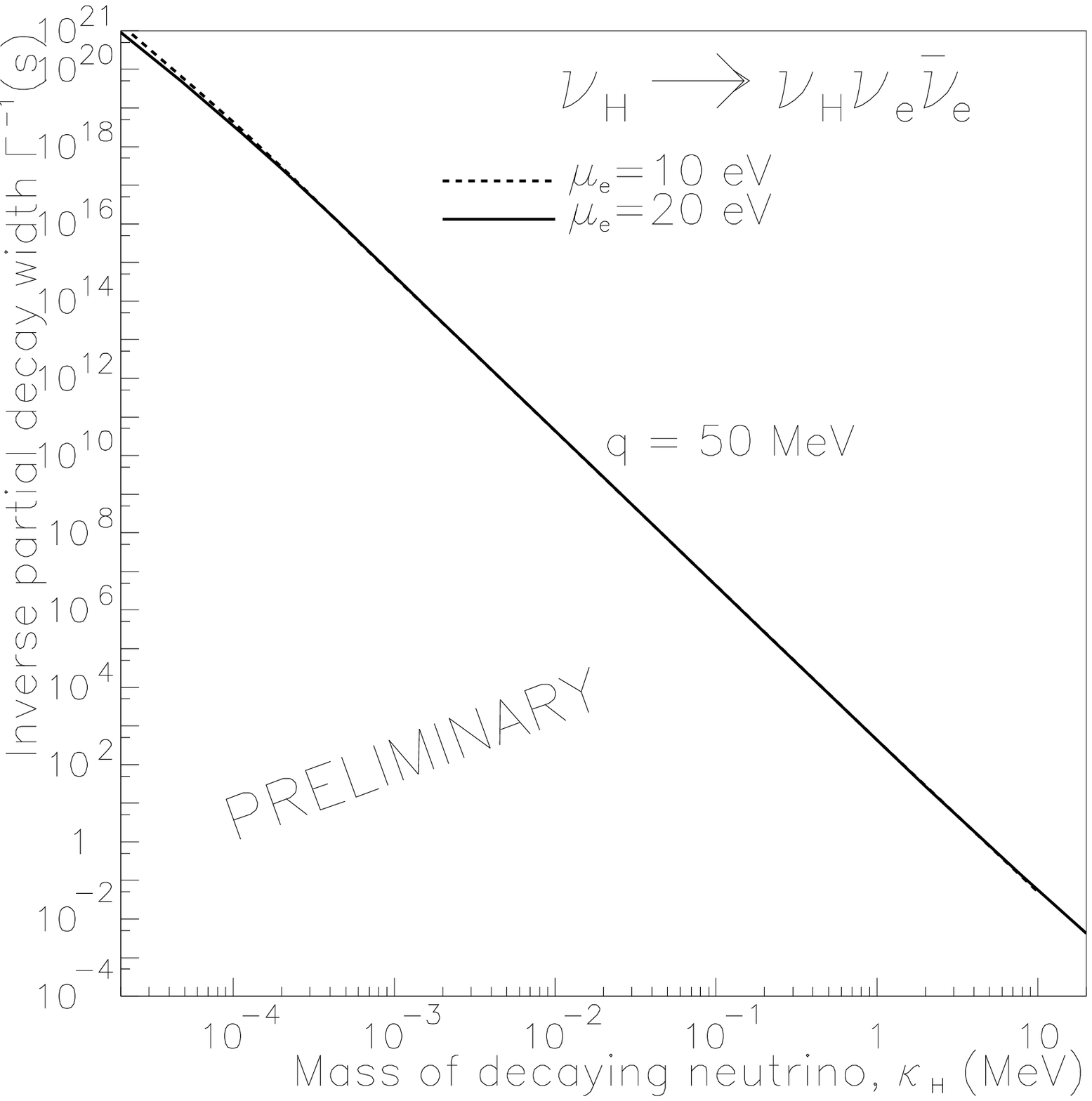}{10cm}{10cm}
  {{\sl  Dependence of the inverse partial decay width on  
the mass of decaying neutrino, $\kappa_{H}$, for the decay into 
electron  neutrino -- antineutrino  pair ($\mu_{e}=10,20\;{\rm eV}$). }}
  {ryc:3b_fig3}

\section{Comments}

In this section we list a number of remarks concerning  experimental
aspects of the hypothesis of tachyonic neutrinos.
The first and the easiest task is  to reanalyse  the existing tritium 
decay data by fitting   eq.~\ref{eq:dGdE1} for the electron energy spectrum.
A positive result of this procedure would not be conclusive for the
question whether neutrinos are tachyons since one can never exclude 
a possible explanation based on conventional physics.
An agreement should be reached concerning the shape of the
experimental electron energy spectrum near the endpoint; results
of experiments differ as to position and width of the enhancement.
It is  desirable to measure  the electron energy spectrum near the endpoint
and search for possible `{\it anomalies}' 
in $\beta$ decay of a nucleus other than tritium. One such experiment,
using  $^{187}{\rm Re}$,  is in progress  (endpoint energy
$2.48\pm0.04\;{\rm keV}$) \cite{187Re}.
Redetermination of  the muon neutrino mass squared with increased precision
would be of great importance but  very difficult experimentally.

The theory of tachyonic  neutrino decays and first results  presented above
open  a new field which  ought to be investigated in detail.
Search for decays of   hypothetical  tachyonic neutrinos  is certainly
an experimental challenge. Most suited are  long baseline
exepriments which are planned in future with a purpose of searching
for neutrino oscillations.  Decay probability increases as a power of 
the beam energy so the highest beam energies are most desirable. 
The results presented above indicate that decays should be
searched for the heavy (muon?) neutrino in the channel: 
$\nu_{\mu}\rightarrow\nu_{\mu}\nu_{\mu}\overline{\nu}_{\mu}$  
(i.e. with experimental signature $\nu_{\mu}\rightarrow\overline{\nu}_{\mu}$) 
which has a larger branching ratio than the one with the 
$\nu_{e}\overline{\nu}_{e}$ pair in the final state.

\section{Summary and conclusions}

According to the results of several  recent high precision experiments,
both electron and muon neutrinos are  found
to  have  negative values of their masses squared.
The significance of this statement  in the case of the electron neutrino
is supported by consistent results obtained in independent measurements. 
The result for the muon neutrino is not conclusive.
Measurements of the Mainz, Troitsk and LLNL groups 
\cite{backe96,lobashev96,stoeffl95} reveal an enhancement  
in the integral electron spectrum  near the endpoint.
We calculated the amplitude for the beta decay of tritium involving
the tachyonic neutrino in the framework of the  causal theory of tachyons 
\cite{jaremb1,jaremb2,jaremb3}.
As a result we reproduce qualitatively
the enhancement  observed in experimental spectra.

We presented results  of preliminary  studies concerning 
three body tachyonic neutrino decays. 
In general the decay probability increases with neutrino energy
according to a power law.
Tachyonic electron neutrinos  are  stable
(mean lifetime  exceeds the age of the Universe) 
in  a wide range of initial neutrino energies.
For the purpose  of making reliable  predictions
more precise measurements of  the muon and tau neutrino masses
are  needed. More studies are in progress.

\vspace{1.0cm}

\noindent
{\bf Acknowledgements}

We wish to thank K.~A. Smoli\'nski and P. Caban for helping us with 
numerical calculations.

\newpage 

\section{Appendix}

The (differential) energy spectrum  of  electrons in $\beta$
decay, $d\Gamma /dl^0$, may be obtained by means of  
formulae (\ref{eq:beta1}),(\ref{eq:beta2}), (\ref{eq:beta3})
after elementary integration of:
\begin{eqnarray}\label{eq:dGdE1}
\frac{d\Gamma}{dl^0}= \frac{1}{4 (2 \pi)^5 m_n}
\int^{r_{+}}_{\max{(r_{-},0)}}dr^0\,
\left|M(l^0,r^0)\right|^2
\end{eqnarray}
with
\begin{eqnarray*}
\lefteqn{r_{\pm}=\frac{-\Delta m^2 l^0 + \Delta m^2 m_n +
2 (l^0)^2 m_n - 2 l^0 m_{n}^{2}}
{2(m_{e}^{2} - 2 l^0 m_n + m_{n}^{2})}+\mbox{}}\\
&&\mbox{}\pm\frac{\sqrt{((l^0)^2 - m_{e}^{2})((\Delta m^2)^2 +
4 \kappa^2 m_{e}^{2} -
4 \Delta m^2 l^0 m_n - 8 \kappa^2 l^0 m_n + 4 \kappa^2 m_{n}^{2} +
4 (l^0)^2 m_{n}^{2}}}{2(m_{e}^{2} - 2 l^0 m_n + m_{n}^{2})}
\end{eqnarray*}
where $\Delta m^2=m_{n}^{2}-m_{p}^{2}+m_{e}^{2}-\kappa^2$
and  the following expression for the matrix element squared:

\begin{eqnarray*}\label{eq:dGdE2}
\lefteqn{\left|M(l^0,r^0)\right|^2=32\,G_{F}^{2}\,m_n\,\pi^2\left\{
\left[2(1-g_{A}^{2})m_e m_p \kappa -
(1+g_{A}^{2})m_e m_n \kappa\right]\right.+\mbox{}}\\
&&\mbox{}+\left[(1+g_{A}^{2})m_e \kappa-2 g_{A} \kappa^2
- g_A \Delta m^2 \right]l^0 + \mbox{}\\
&&\mbox{}+\left[(1+g_{A}^{2})m_e \kappa-2 g_{A} m_{e}^{2}
+ g_A \Delta m^2 \right]r^0 + \mbox{}\\
&&\mbox{}-2 g_A m_n (r^0)^2 + 2 g_A m_n (l^0)^2 + \mbox{}\\
&&\mbox{}+\frac{1}{\sqrt{(r^0)^2+\kappa^2}}\left\{
\left[(1+g_{A}^{2})\kappa^2\left(\frac{1}{2}\Delta m^2-m_{e}^{2}\right)+
g_A m_e \kappa(2\kappa^2+\Delta m^2)\right]\right.+\mbox{}\\
&&\mbox{}\quad+\left[(1+g_{A}^{2})m_n\kappa^2-
(1-g_{A}^{2})m_p\kappa^2-2 g_A m_e m_n \kappa\right]l^0+\mbox{}\\
&&\mbox{}\quad+\left[\frac{1}{2}(1-g_{A}^{2}) m_p \Delta m^2 -
(1+g_{A}^{2}) m_n \kappa^2 - 2 g_A m_e m_n \kappa\right]r^0+\mbox{}\\
&&\mbox{}\quad+\left[\frac{1}{2}(1+g_{A}^{2})\Delta m^2 +
2 g_A m_e \kappa - (1-g_{A}^{2})m_n m_p\right]l^0 r^0+\mbox{}\\
&&\mbox{}\quad+\left[2 g_A m_e \kappa - (1-g_{A}^{2})m_n m_p+
(1+g_{A}^{2})\left(\frac{1}{2}\Delta m^2 - m_{e}^{2}\right)
\right](r^0)^2+\mbox{}\\
&&\mbox{}\quad-\left.\left.(1+g_{A}^{2})\kappa^2(l^0)^2 -
(1+g_{A}^{2})m_n r^0 (l^0)^2 - (1+g_{A}^{2})m_n (r^0)^3\right\}\right\}.
\end{eqnarray*}



\end{document}